\documentclass{aa}  

\usepackage{graphicx}
\usepackage{txfonts}
\newcommand{\numat}{\mbox{\boldmath $\nu$}}

\newcommand{\Smat}{\mbox{\boldmath $\cal S$}}

\newcommand{\Xmat}{\mbox{\boldmath $X$}}

\begin{document}

   \title{Electron impact ro-vibrational transitions and dissociative recombination of H$_2^+$ and HD$^{+}$}

   \subtitle{Rate coefficients and astrophysical implications}

   \author{R. Hassaine\inst{1,*},
          E. Djuissi\inst{1},
          N. Pop\inst{2},
          F. Iacob\inst{3},
          M.~D. Epée Epée\inst{4},
          O. Motapon\inst{4},
          V. Laporta\inst{5},
          R. Bogdan\inst{6},
          M. Ayouz\inst{7},
          M. Telmini\inst{8},
          C.~M. Coppola\inst{9},
          D. Galli\inst{10},
          J. Zs. Mezei\inst{1,11},
          \and I.~F. Schneider\inst{1,12}
          }

   \institute{Laboratoire Ondes et Milieux Complexes, UMR 6294 CNRS and Universit\'e  Le Havre Normandie, 25 rue Philippe Lebon, BP 540, 76058, Le Havre, France\\
              *\email{riyad.hassaine@univ-lehavre.fr}
         \and
             Department of Physical Foundation of Engineering, Politehnica University of Timi\c{s}oara, Blvd. Vasile P\^{a}rvan 2, 300223 Timi\c{s}oara, Romania
         \and
             Dept. of Physics, West University of Timi\c{s}oara, Blvd. Vasile P\^{a}rvan 4, \\
             300223 Timi\c{s}oara, Romania
         \and
             LPF, UFD Math\'ematiques, Informatique Appliqu\'ee et Physique Fondamentale, University of Douala, P. O. Box, \\ 24157, Douala, Cameroon
         \and
             Instituto per la Scienza e Tecnologia dei Plasmi, CNR, Bari, Italy
         \and
             Department of Comput. Informat. Technol. Engineering, Politehnica University of Timi\c{s}oara, Bv. Vasile P\^{a}rvan 2, \\ 300223 Timi\c{s}oara, Romania
         \and
             LGPM, CentraleSup\'elec, Universit\'e Paris-Saclay, 8-10 Rue Joliot Curie, F-91190 Gif-sur-Yvette, France
         \and
             LSAMA, Universit\'e de Tunis El Manar, Tunis, Tunisia
         \and
             Department of Chemistry, Università degli Studi di Bari Aldo Moro, 70125 Bari, Italy
         \and
             INAF Osservatorio Astrofisico di Arcetri, 50125 Firenze, Italy
         \and
             HUN-REN Institute for Nuclear Research (ATOMKI), Bem Sqr. 18/c, 4026 Debrecen, Hungary
         \and
             LAC CNRS-FRE2038, Universit\'e Paris-Saclay, ENS Cachan, Campus d'Orsay, Bat. 305, 91405 Orsay, France
             }

   \date{July 19, 2025}

 
  \abstract
  {Molecular hydrogen and its cation H$_2^+$ are among the first species formed in the early Universe, and play a key role in the thermal and chemical evolution of the primordial gas. In molecular clouds, H$_2^+$ ions formed through ionization of H$_2$ by particles react rapidly with H$_2$ to form H$_3^+$, triggering the formation of almost all detected interstellar molecules.}
  {We present a new set of cross sections and rate coefficients for state-to-state ro-vibrational transitions of the H$_2^+$ and HD$^+$ ions, induced by low-energy electron collisions. Study includes the major electron-impact processes relevant for low-metallicity astrochemistry: inelastic and superelastic scattering, and dissociative recombination.}
  {The electron-induced processes involving H$_2^+$ and HD$^+$ are treated using the multichannel quantum defect theory.}
  {The newly calculated thermal rate coefficients show significant differences compared to those used in previous studies. When introduced into astrochemical models, particularly for shock-induced chemistry in metal-free gas, the updated dissociative recombination rates produce substantial changes in the predicted molecular abundances.}
  {These data provide updated and improved input for the modeling of hydrogen-rich plasmas in environments where a high abundance of free electrons is expected, such as planetary nebulae, H{\sc ii} regions, and the ionospheres of giant planets.}

\titlerunning{Electron impact transitions of H$_2^+$ and HD$^+$}
\authorrunning{R. Hassaine et al.}

\maketitle

%

\section{Introduction}

Molecular hydrogen and its cation are among the first species formed in the early Universe \citep[for a review, see, e.g.][]{galli2013}. Due to the lack of a dipole moment, in the absence of catalyzing dust grains, the homonuclear molecule H$_2$ cannot be formed by radiative association of two hydrogen atoms, and its production in a gas of primordial composition proceeds via the intermediary species H$^-$ and H$_2^+$ \citep[see, e.g.,][]{lepp2002,coppola2011,gay2012}. The latter
channel is the dominant one at high temperature, and is 
limited by the destruction of H$_2^+$ by photodissociation, charge exchange with H atoms, and dissociative recombination (DR) with electrons.
In dense molecular clouds where ultraviolet radiation is efficiently absorbed by dust grains, H$_2^+$ is formed in excited vibrational states by cosmic-ray ionization of 
H$_2$ \citep{glassgold1973} and rapidly transformed into H$_3^+$, 
which is the starting
point of gas phase ion-molecule reaction chemistry \citep[see, e.g.,][]{tielens2013}.
In addition, due to the high electron densities, DR of H$_2^+$ 
plays an important role also in brown dwarfs atmospheres \citep{gibbs2022, pineda2024}, planetary nebulae \citep{black1978}, and H{\sc ii} regions \citep{aleman2004}. 

The molecular hydrogen cation is also the primary molecular ion formed by photoionization of H$_2$ in the upper atmospheres and ionospheres of outer gaseous planets in the Solar System and extrasolar giant planets \citep[see, e.g.][]{chadney2016}. In this context, the recombination of H$_2^+$ with electrons has also been studied as a source of excited states of atomic hydrogen, in particular H($2p$), in models of the ultraviolet emission of Jupiter's upper atmosphere, where it competes with the energy transfer between free protons and H($2s$) \citep{shem1985}.

Dissociative recombination is also an important reaction in laboratory astrophysics~\citep{Pari} because it affects the diagnostics of electric discharges in hydrogen and deuterium mixtures based on the line profile of the hydrogen Balmer series, where this recombination process becomes dominant. The effect has been studied in particular in recombining plasmas where the electron temperature is below 1~eV \citep{Frie,Wund} and therefore relevant for the interstellar medium. The effect of dissociative recombination on diagnostics was also studied using the Monte Carlo DEGAS2 model \citep{stotler1994}, applied to the simulation of MAP-II (materials and plasma) columnar plasma sources \citep{Tana}, used to study edge recombination. In this type of source, the plasma has two mixed but not balanced components: one has a temperature of the order of 1~eV, and a second has a higher temperature. \cite{Kras} used a collisional-radiative model to show that, due to this reaction too, the formation of molecular ions has a significant effect on the recombination rate of a hydrogen plasma.

In all these environments, in which the ionization fraction is significant, the electron-impact induced ro-vibrational transitions (RVT), strong competitors of the DR, provide the dominant path to excitation and de-excitation process of H$_2^+$ \citep{black1987, lepetit2002, roueff2010}.
The aim of the present paper is to provide accurate cross sections and rate coefficients for electron-impact induced RVT,
\begin{equation}
\label{eq:RVT}
\text{H$_2^+$}(N_{i}^{+}, v_{i}^{+}) + e(\varepsilon) \longrightarrow \text{H$_2^+$}(N_{f}^{+}, v_{f}^{+}) + e({\varepsilon^\prime}),
\end{equation}
and DR reactions,  
\begin{equation}
\label{eq:DR} 
\text{H$_2^+$}(N_{i}^{+}, v_{i}^{+}) + e(\varepsilon) \longrightarrow \text{H}(1s) + \text{H}(n\ge 2),
\end{equation}
of H$_2^+$ and its isotopologue HD$^+$, including all relevant ro-vibrational levels. In reactions~(\ref{eq:RVT}) and (\ref{eq:DR}), $N_i^+/N_f^+$ and $v_i^+/v_f^+$ are the initial/final rotational and vibrational quantum numbers of the target ion, respectively, $\varepsilon/\varepsilon^\prime$ is the energy of the incident/scattered electron, and $n$ the principal quantum number of the excited atom resulting from dissociation. 
RVT can be further characterized as resonant elastic collisions, inelastic collisions and super-elastic collisions when the energy 
$\varepsilon^\prime$ of the scattered electron is equal to, smaller, or larger than the energy $\varepsilon$ of the incident electron.
The spectroscopic information, given by the output of reactions (\ref{eq:RVT})--(\ref{eq:DR}), is sensitive to the initial $(N_i^+$, $v_i^+)$ and final ($N_f^+$, $v_f^+$) quantum numbers of the target ions and consequently provides the structure of the ionized media.


This study was initiated in previous publications by our group \citep{motapon2014, epee2016, djuissi2019, epee2022}.
We notice that, whereas for H$_2^+$ the rotational transitions involve rotational quantum numbers of strictly the same parity  ({\it gerade} or {\it ungerade}), this rule is not valid for the deuterated variant. However, we have assumed it for HD$^+$ too, since no data are currently available for the {\it gerade/ungerade} mixing, and since the transitions between rotational quantum numbers of different parity are much less intense than the others \citep{shafir2009}. 

This paper is organized as follows: In Sect.~\ref{sec:theory} we describe our theoretical method. In Sect.~\ref{sec:res} we present our results in terms of cross sections and rate coefficients. In Sect.~\ref{sec:app} we discuss the applications in the media of astro-chemical interest. And, finally, in Sect.~\ref{sec:con} we 
summarize our conclusions.

\section{Theoretical Method}
\label{sec:theory}

\begin{table*}
	\centering
	\caption{The energy of the first 30 ro-vibrational levels of H$^+_2$ and HD$^+$ $X^2\Sigma_g^+$ electronic states relative to the ground $(N_i^+,v_i^+)=(0,0)$ level.}
	\label{tab:1}
	\begin{tabular}{ccc@{\extracolsep{5pt}}cc@{\extracolsep{5pt}}ccc@{\extracolsep{5pt}}cc} 
		\hline
		& \multicolumn{2}{c}{H$_2^+$}& \multicolumn{2}{c}{HD$^+$} & &\multicolumn{2}{c}{H$_2^+$}& \multicolumn{2}{c}{HD$^+$}\\
		\cline{2-3}\cline{4-5}
		\cline{7-8}\cline{9-10}
		 no & $(N_{i}^{+},v_{i}^{+})$ & energy (eV) & $(N_{i}^{+},v_{i}^{+})$ & energy (eV)& no & $(N_{i}^{+},v_{i}^{+})$ & energy (eV) & $(N_{i}^{+},v_{i}^{+})$ & energy (eV) \\
		\hline
		1 & (0,0) & 0.000 &(0,0)& 0.000&16 &(10,0)&0.371 &(4,1) & 0.289\\
		2 & (1,0) & 0.007& (1,0)& 0.005 & 17&(5,1) & 0.372& (5,1)& 0.314 \\
		3 & (2,0) &0.021 & (2,0) &0.016 & 18&(6,1)&0.412 & (11,0) &0.337 \\
		4 &(3,0) & 0.043 & (3,0) & 0.033 & 19 & (11, 0)& 0.44&(6,1)&0.344\\
		5 & (4,0)& 0.071&(4,0)&0.054 &20&(7,1)&0.457&(7,1) & 0.379 \\
		6 & (5,0)&0.106&(5,0)& 0.080 & 21&(8,1)& 0.507& (12,0) & 0.394 \\
		7 &(6,0)&0.148&(6,0) & 0.112 & 22&(12,0) &0.512& (8,1) & 0.418 \\
		8 & (7,0)&0.195&(7,0) & 0.148 & 23 &(0,2)&0.528&(13,0) & 0.455 \\
		9 & (8,0) &0.249&(8,0) & 0.189 & 24 & (1,2)&0.534&(9,1) & 0.461 \\
		10 &(0,1)&0.272& (9,0) & 0.234 & 25 &(2,2)&0.547& (0,2) & 0.462 \\
		11 & (1,1)&0.279&(0,1) & 0.237 & 26 & (9,1)&0.563&(1,2) & 0.467 \\
		12 & (2,1)&0.292&(1,1) & 0.242 & 27 &(3,2)&0.566& (2,2) & 0.477 \\
		13 &(9,0)&0.307& (2,1) & 0.253 & 28&(13,0)& 0.589& (3,2) & 0.492 \\
		14 &(3,1)&0.312&(3,1) & 0.268 & 29&(4,2)&0.592 & (10,1) & 0.508 \\
		15 &(4,1)&0.339& (10,0) & 0.284 & 30 &(5,2)&0.623& (4,2) & 0.511 \\
		\hline
	\end{tabular}
\end{table*} 

The electron-induced reactions (\ref{eq:RVT}) and (\ref{eq:DR}) for H$_2^+$ and HD$^+$ are studied by a step-wise version of the multichannel quantum defect theory (MQDT), which we successfully applied to calculate dissociative recombination, ro-vibrational and dissociative excitation cross sections for molecular cations such as H$_2^+$ and its isotopologues~\citep{waffeu2011,chakrabarti2013,motapon2014, epee2016, epee2022}, CH$^+$~\citep{mezei2019}, SH$^+$~\citep{kashinski2017}, BeH$^+$ and its isotopologues~\citep{niyonzima2017, niyonzima2018, pop2021}.

A detailed description of the method has been given in previous articles~\citep{motapon2014,mezei2019}, and we list below only
the main steps of our current MQDT treatment:
\begin{enumerate}
\item {\it Building the interaction matrix}  based on the couplings between ionization channels - associated with the ro-vibrational levels $N^+$, $v^+$ of the cation and with the orbital quantum number $l$ of the incident/Rydberg electron - and the dissociation channels.
\item {\it Computation of the reaction matrix} $K$ as a second-order perturbative solution of the Lippmann-Schwinger
integral equation
~\citep{ngassam03b}.
\item {\it Diagonalization of the reaction matrix} by building the short-range representation of the eigenchannel.
\item {\it Frame transformation} from the Born-Oppenheimer (short-range) representation—characterized by quantum numbers $N$, $v$, $\Lambda$, and $l$—to the close-coupling (long-range) representation, defined by $N^+$, $v^+$, and $l$. Here, $N$ and $N^+$ denote the total rotational quantum numbers of the neutral system and the ion, respectively, $v$ and $v^+$ are the corresponding vibrational quantum numbers, and $l$ is the orbital angular momentum of the incoming, outgoing, or scattered electron.
\item {\it Building of the generalized scattering matrix} based on the frame-transformation coefficients 
organized in blocks associated with energetically open ($o$) and/or closed ($c$) channels:
\begin{equation}
\Xmat=
\left(\begin{array}{cc} \Xmat_{oo} & \Xmat_{oc}\\
\Xmat_{co} & \Xmat_{cc} \end{array} \right).
\end{equation}

\item {\it  Building of the physical scattering matrix}:
\begin{equation}\label{eq:solve3}
\Smat=\Xmat_{oo}-\Xmat_{oc}\frac{1}{\Xmat_{cc}-\exp(-2i\pi \numat)}\Xmat_{co}.
\end{equation}
obtained by the so-called ``elimination of closed channel'' \citep{seaton1983}.
The diagonal matrix {\numat} in the denominator
contains 
the effective quantum numbers corresponding to the vibrational thresholds of the closed ionization channels at the current total energy of the system. 

\item{\it Computation of the cross-sections} for the target initially in a state characterized  by the quantum numbers $N_i^+,v_i^+$, and for the energy of the incident electron $\varepsilon$, the ro-vibrational transitions and the dissociative recombination global cross-sections read respectively 
\begin{eqnarray}
& & \sigma _{N_{f}^{+}v_{f}^{+} \leftarrow N_{i}^{+}v_{i}^{+}}  = 
\sum_{\Lambda,{\rm sym}} \sigma _{N_{f}^{+}v_{f}^{+}\leftarrow N_{i}^{+}v_{i}^{+}}^{({\rm sym},\Lambda)},
\label{eqVE_VdE1} \\
& & \sigma _{N_{f}^{+}v_{f}^{+}\leftarrow N_{i}^{+}v_{i}^{+}}^{({\rm sym},\Lambda)}  = 
\frac{\pi}{4\varepsilon} \rho^{({\rm sym},\Lambda)} \sum_{N} \frac{2N+1}{2N_{i}^{+}+1} \nonumber \\
& & \times \sum_{l,l'}\mid S_{N_{f}^{+} v_{f}^{+}l',N_{i}^{+} v_{i}^{+}l}^{({\rm sym},\Lambda,N)} - \delta_{N_{i}^{+}N_{f}^{+}}\delta_{v_{i}^{+}v_{f}^{+}}\delta_{l'l}\mid^2, 
\label{eqVE_VdE2} \\
& & \sigma _{{\rm diss} \leftarrow N_{i}^{+}v_{i}^{+}} = 
\sum_{\Lambda,{\rm sym}} \sigma_{{\rm diss} \leftarrow N_{i}^{+}v_{i}^{+}}^{({\rm sym},\Lambda)},  \label{eqDR1} \\
& & \sigma_{{\rm diss} \leftarrow N_{i}^{+}v_{i}^{+}}^{({\rm sym},\Lambda)} =  
\frac{\pi}{4\varepsilon} \rho^{({\rm sym},\Lambda)} \sum_{N} \frac{2N+1}{2N_{i}^{+}+1} \nonumber \\
& & \times \sum_{l,j}\mid S^{({\rm sym},\Lambda,N)}_{d_{j},N_{i}^{+}v_{i}^{+}l}\mid^2
\label{eqDR2}
\end{eqnarray}
where $sym$ refers to the inversion symmetry  - {\it gerade/ungerade} - and to the spin quantum number of the neutral system and $\rho^{({\rm sym},\Lambda)}$ is the ratio between the spin multiplicities of the neutral system and that of the ion.
\end{enumerate}

This work extends our previous studies performed on HD$^+$~\citep{waffeu2011,motapon2014} and H$^+_2$~\citep{epee2016} for low collision energies relevant for kinetics modeling in astrochemistry, by considering simultaneous rotational and vibrational transitions (excitations and/or de-excitations) and by increasing the range of the incident energy of the electron. We have used the same molecular structure data sets as those from our previous studies ~\citep{waffeu2011, motapon2014, epee2016, djuissi2019, epee2022}.  
The energies of the first 30 ro-vibrational levels of the $X^2\Sigma_g^+$ electronic state of H$_2^+$ and HD$^+$ relative to the ground level $(N_i^+,v_i^+)=(0,0)$ are listed in Table~\ref{tab:1}.
The molecular states of the neutral (H$_2$, HD) systems taken into account were of $^1\Sigma_{g}^{+}$, $^1\Pi_{g}$,
$^1\Delta_{g}$, $^3\Sigma_{g}^{+}$, $^3\Pi_{g}$,
$^3\Delta_{g}$, $^3\Sigma_{u}^{+}$, and $^3\Pi_{u}$ symmetries.
Consequently, the partial waves considered for the incident electron were
$s$ and $d$ for the $^1\Sigma_{g}^{+}$ states,
$d$ for  $^1\Pi_{g}$, $^1\Delta_{g}$, $^3\Sigma_{g}^{+}$, $^3\Pi_{g}$
and $^3\Delta_{g}$, and $p$ for $^3\Sigma_{u}^{+}$, and $^3\Pi_{u}$.


\section{Results and discussions}\label{sec:res} 

We have computed the RVT cross sections for the lowest 30 ro-vibrational levels of H$_2^+$ and HD$^+$ in their ground $^2\Sigma_g^+$ electronic state, and for collision energies ranging between 
$10^{-5} - 1.09$ eV - this latter value corresponding to the opening of the following dissociation channel - with an energy step of $0.01$ meV. We considered both {\it direct} and {\it indirect} mechanisms - i.e. including {\it open} and {\it closed} channels respectively -  the reaction matrix being evaluated in the second order of the perturbation theory.

Thermally averaged rate coefficients were obtained by convolution of the cross sections with the isotropic Maxwell distribution function of the kinetic energy of the incident electrons:
\begin{equation}\label{rate}
k(T)=\frac{2}{k_{\rm B}T}\sqrt{\frac{2}{{\pi mk_{\rm B}T}}}\int_{0}^{+\infty}\sigma(\varepsilon)\varepsilon\exp(-\varepsilon/k_{\rm B}T)\,d\varepsilon,
\end{equation}
where $\sigma(\varepsilon)$ are  the cross sections calculated according to eqs.~(\ref{eqVE_VdE1})--(\ref{eqDR2}) and $k_{\rm B}$ is the Boltzmann constant.

\begin{figure*}
\centering
\includegraphics[width=0.46\linewidth]{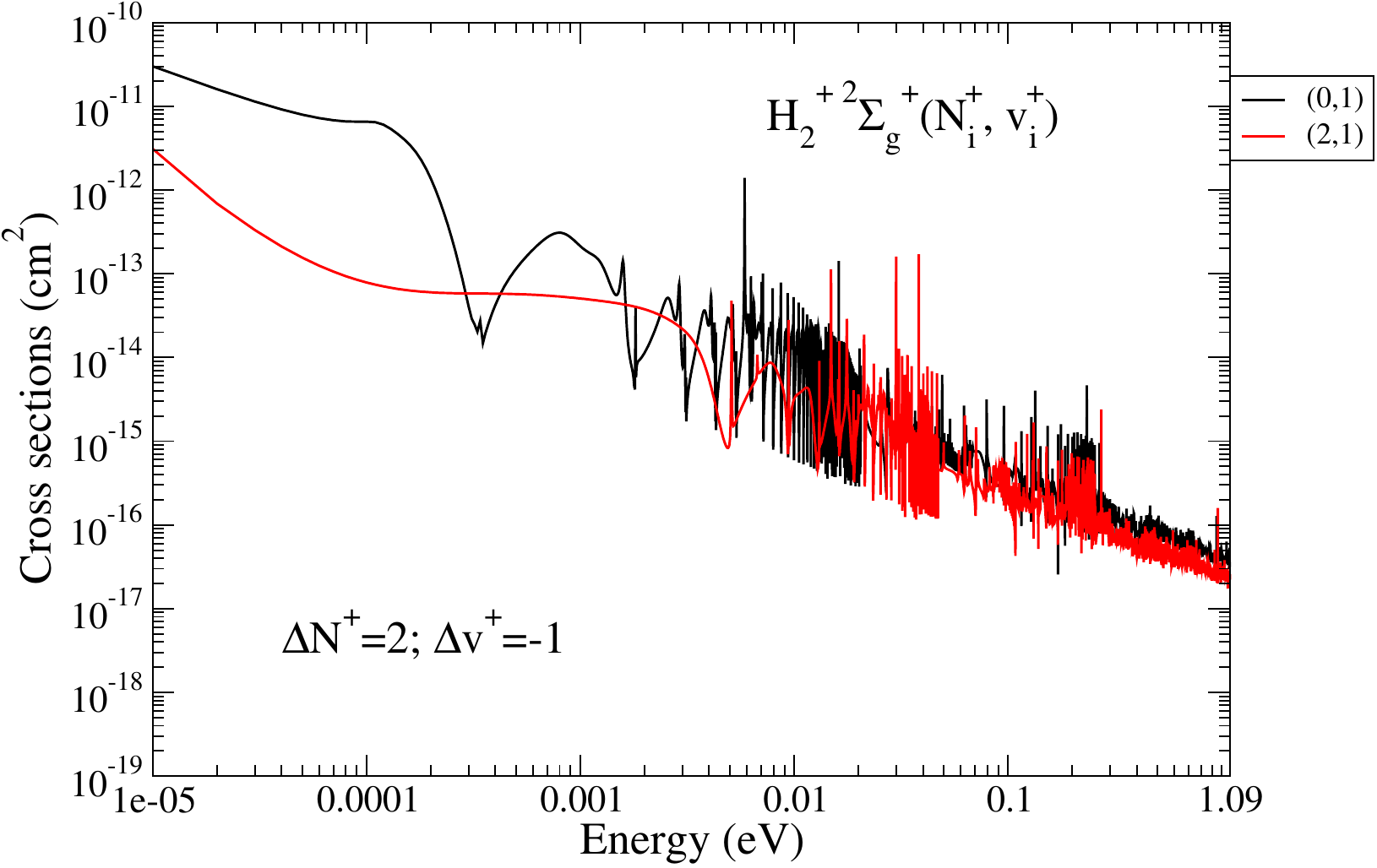}
\includegraphics[width=0.46\linewidth]{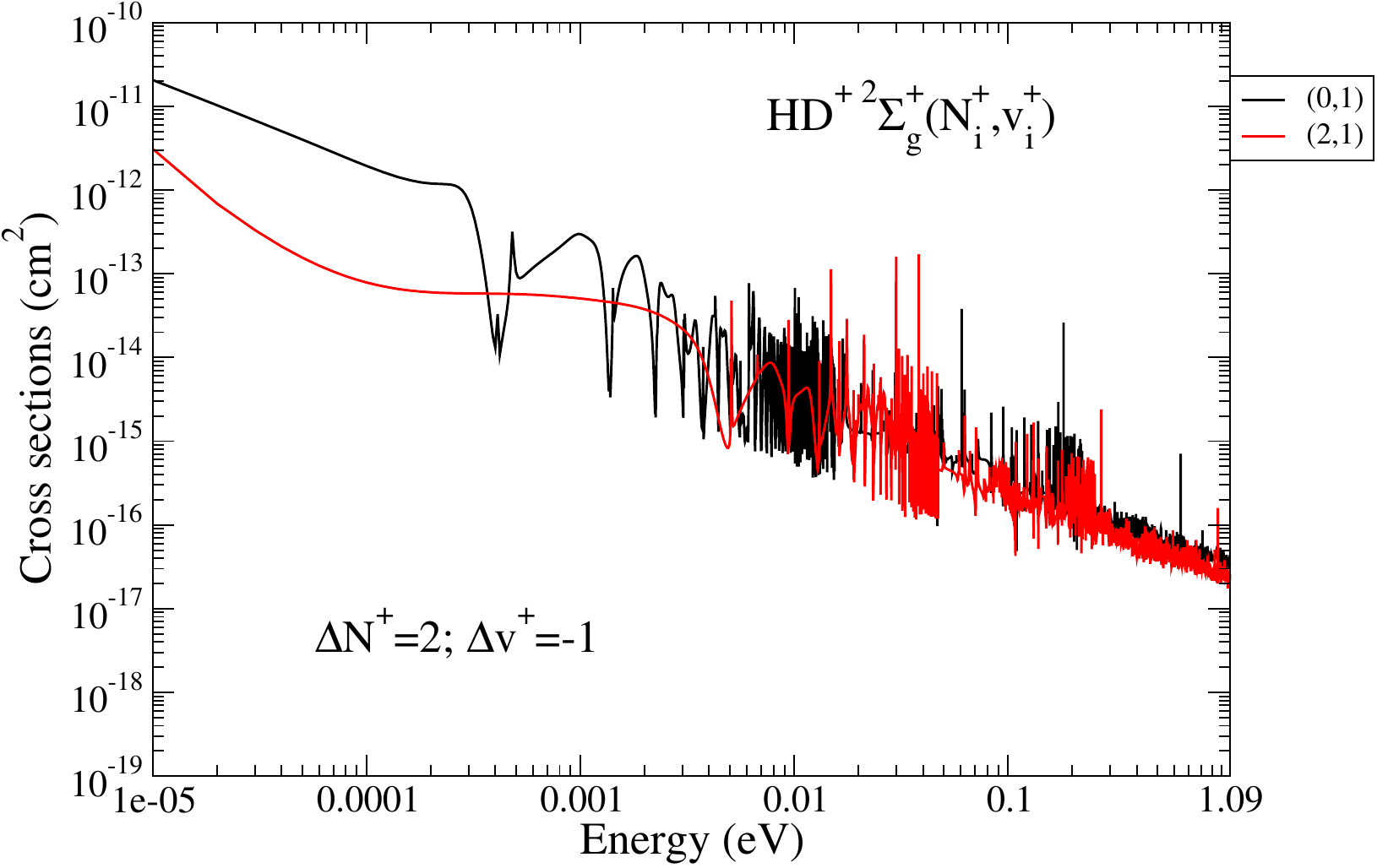}
 \centering
 \caption{{\it Left panel:} 
electron-impact rotational excitation ($\Delta N^+=N_f^+-N_i^+=2$) and vibrational de-excitation ($\Delta v^+=v_f^+-v_i^+=-1$) cross sections of H$_2^+$ from its ground $^2\Sigma_g^+$ electronic state and initial ro-vibrational levels $(N_i^+, v_i^+)=(0,1)$, {\it black line}, and $(N_i^+, v_i^+)=(2,1)$, {\it red line}. {\it Right panel:} same for HD$^+$.}
\label{fig:1}
\end{figure*}

\begin{figure*}
\centering
\includegraphics[width=0.46\linewidth]{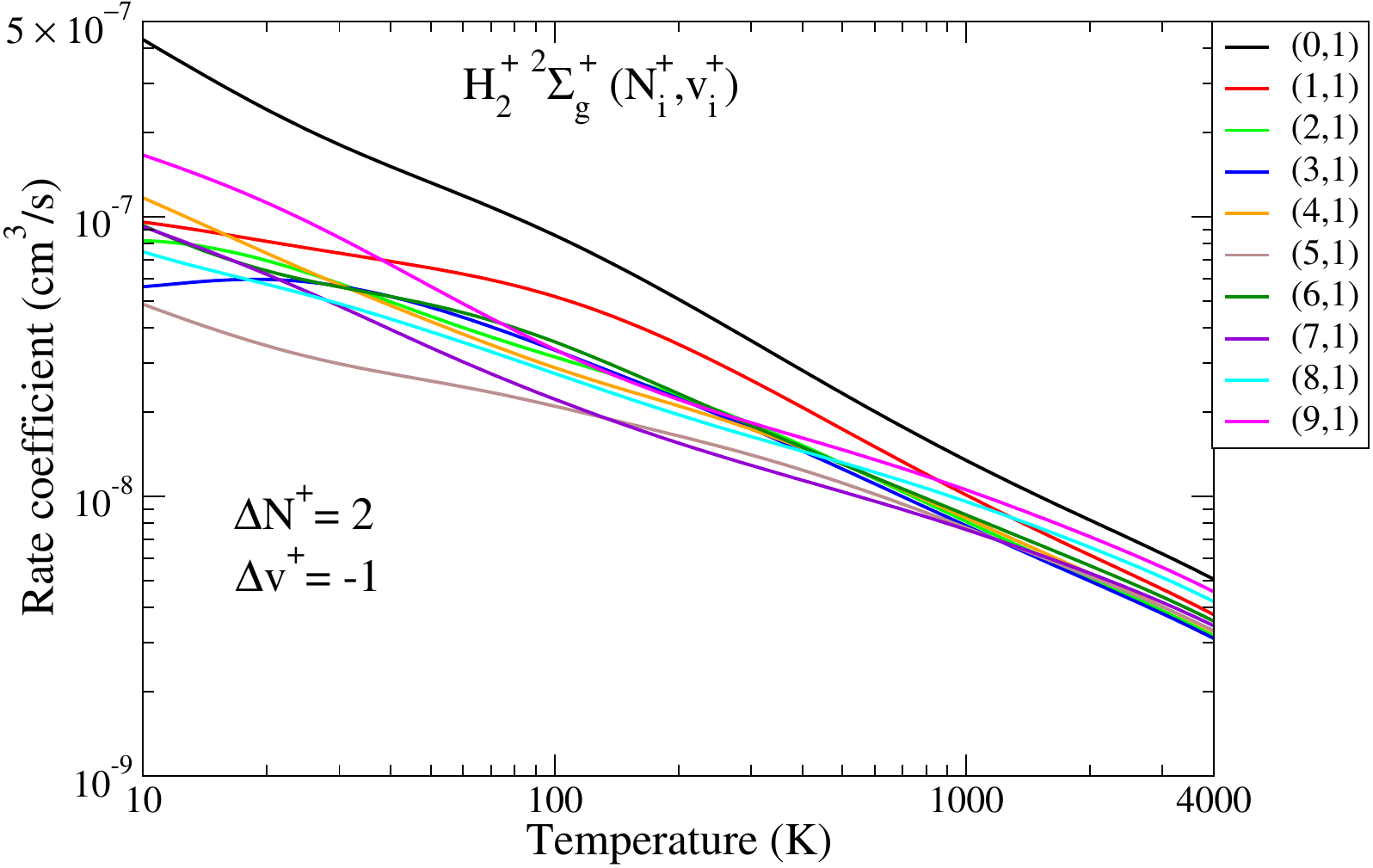}
\includegraphics[width=0.46\linewidth]{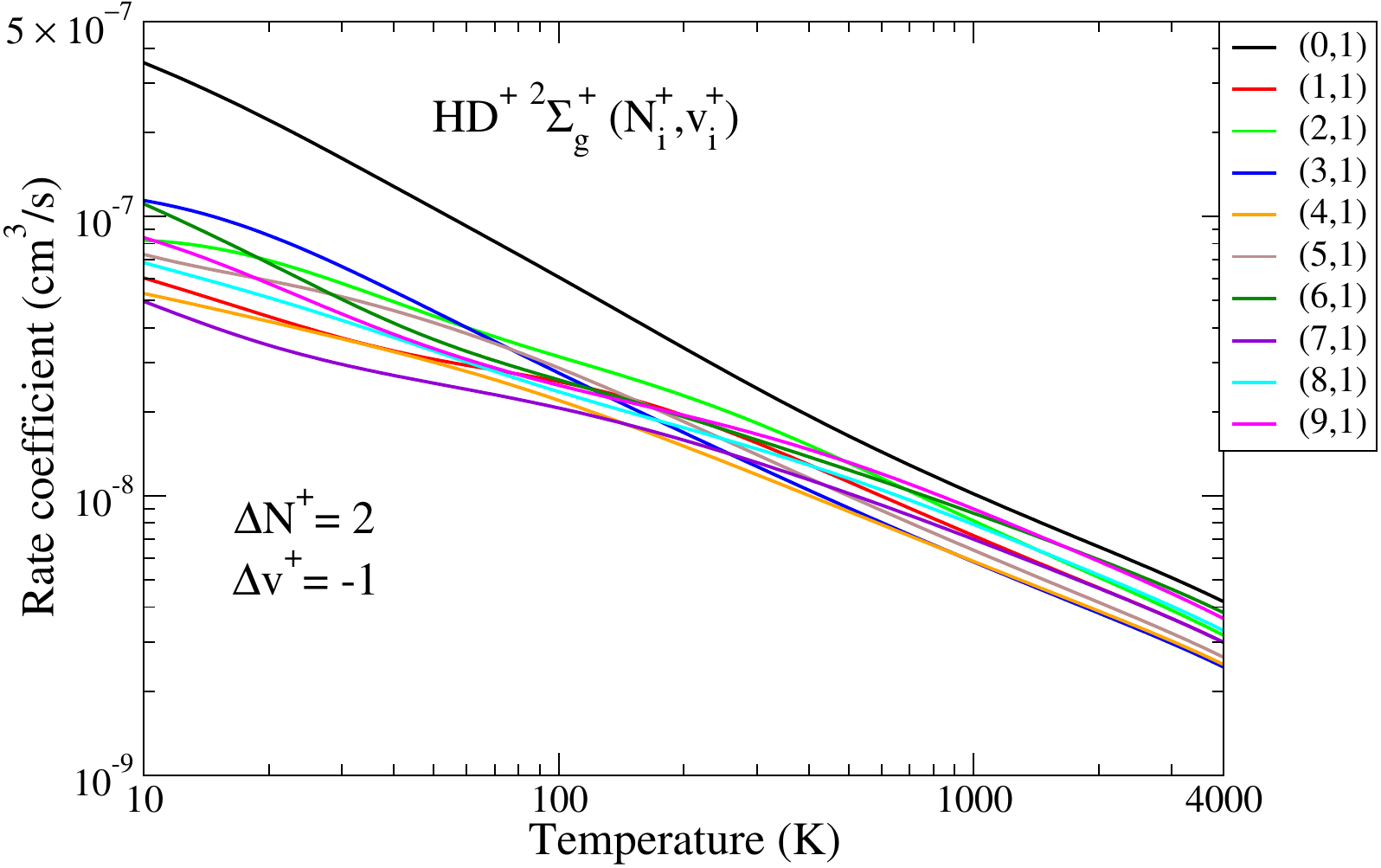}
 \centering
 \caption{{\it Left panel:} electron-impact rotational excitation ($\Delta N^+=N_f^+-N_i^+=2$) and vibrational de-excitation ($\Delta v^+=v_f^+-v_i^+=-1$) rate coefficients of H$_2^+$ from its ground $^2\Sigma_g^+$ electronic state for 10 initial ro-vibrational levels $(N_i^+, v_i^+)$, where $v_i^+=1$. 
 {\it Right panels:} same for HD$^+$.}
\label{fig:2}
\end{figure*}

\begin{figure*}
\centering
\includegraphics[width=0.46\linewidth]{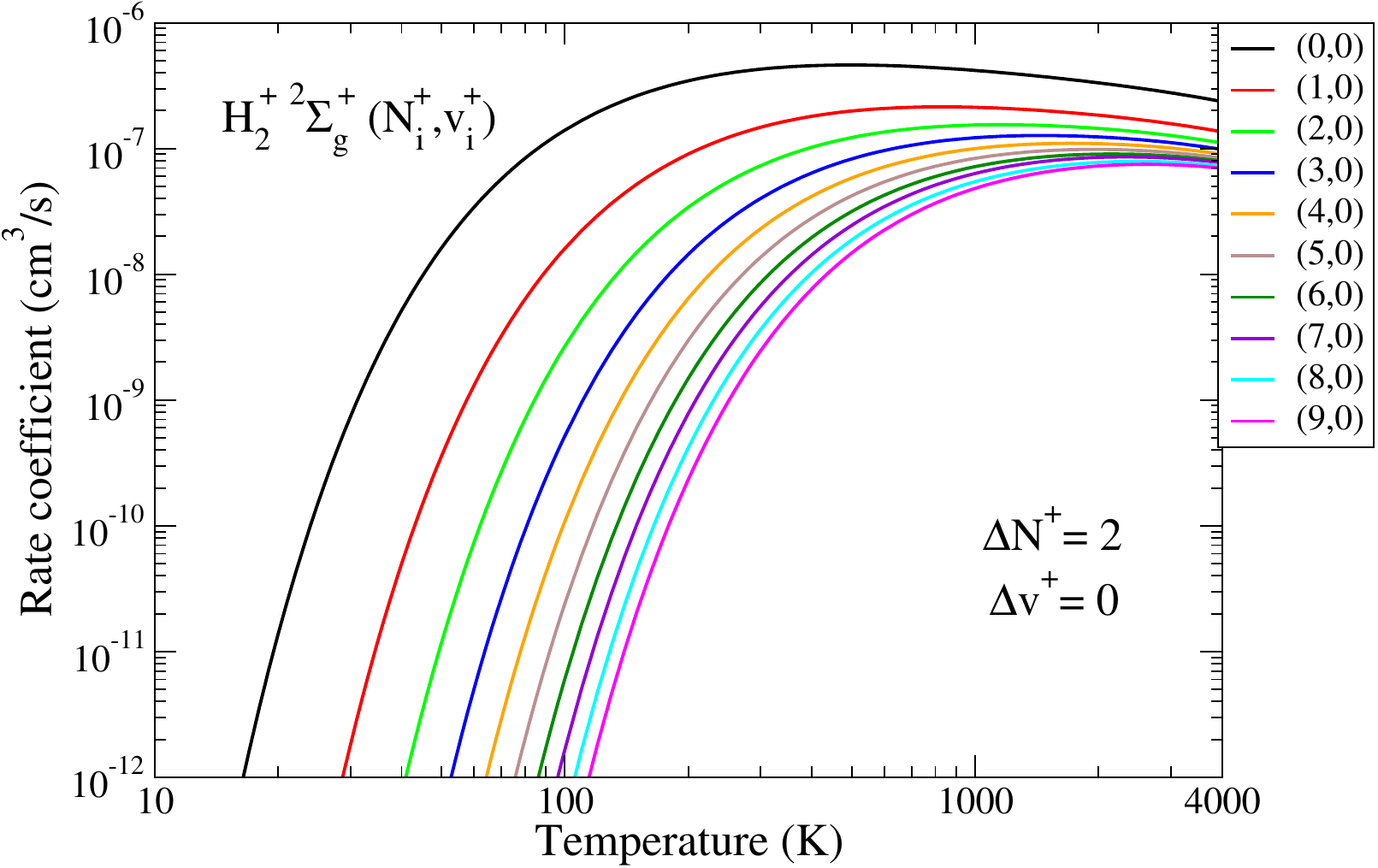}
\includegraphics[width=0.46\linewidth]{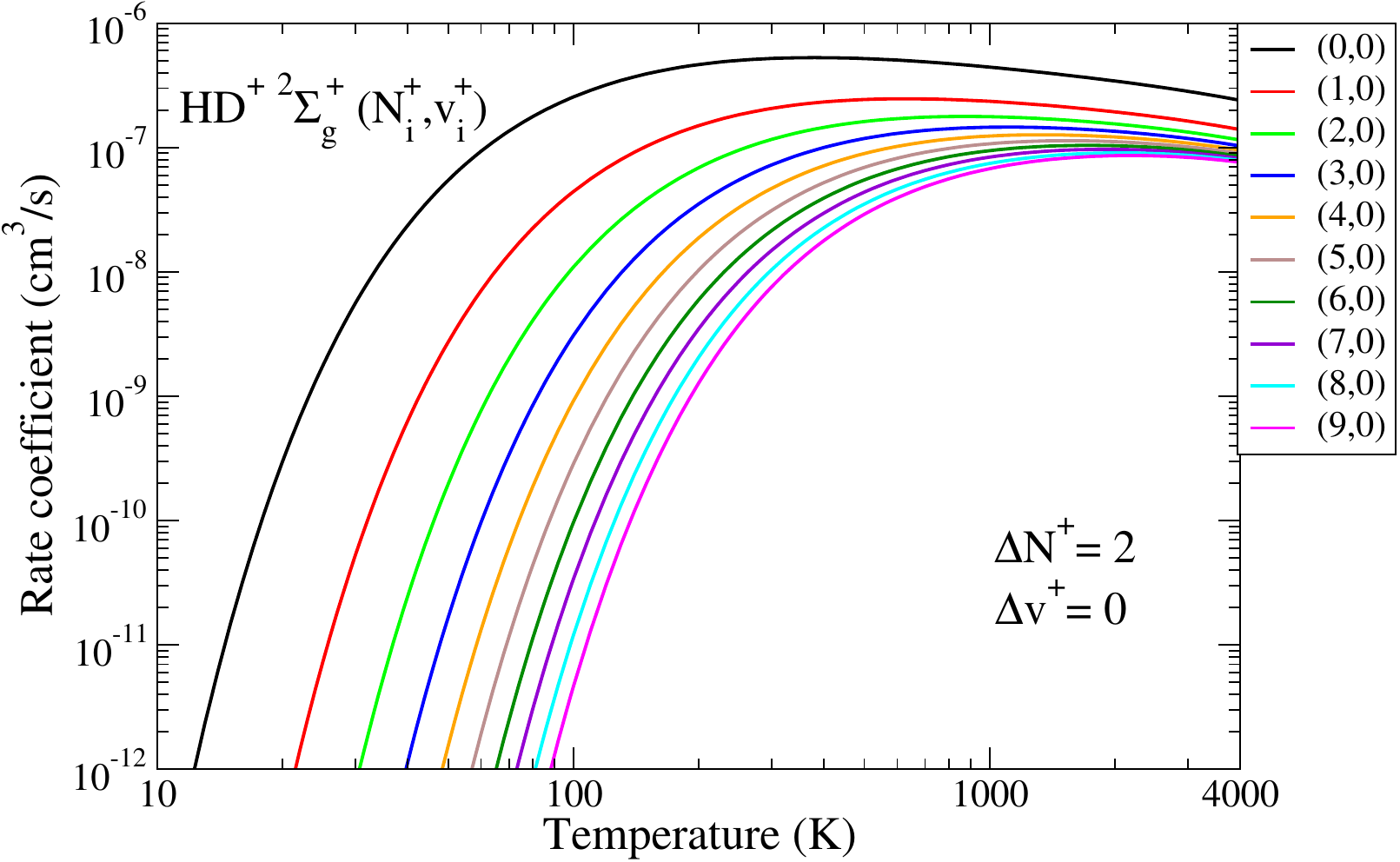}
 \centering
 \caption{{\it Left panel:} 
 electron-impact rotational excitation ($\Delta N^+=N_f^+-N_i^+=2$), ($\Delta v^+=v_f^+-v_i^+=0$) rate coefficients of H$_2^+$ from its ground $^2\Sigma_g^+$ electronic state and for the first 10 initial ro-vibrational levels $(N_i^+, v_i^+)$, where $v_i^+=0$. 
 {\it Right panels:} same for HD$^+$.}
\label{fig:3}
\end{figure*}

\begin{figure*}
\centering
\includegraphics[width=0.46\linewidth]{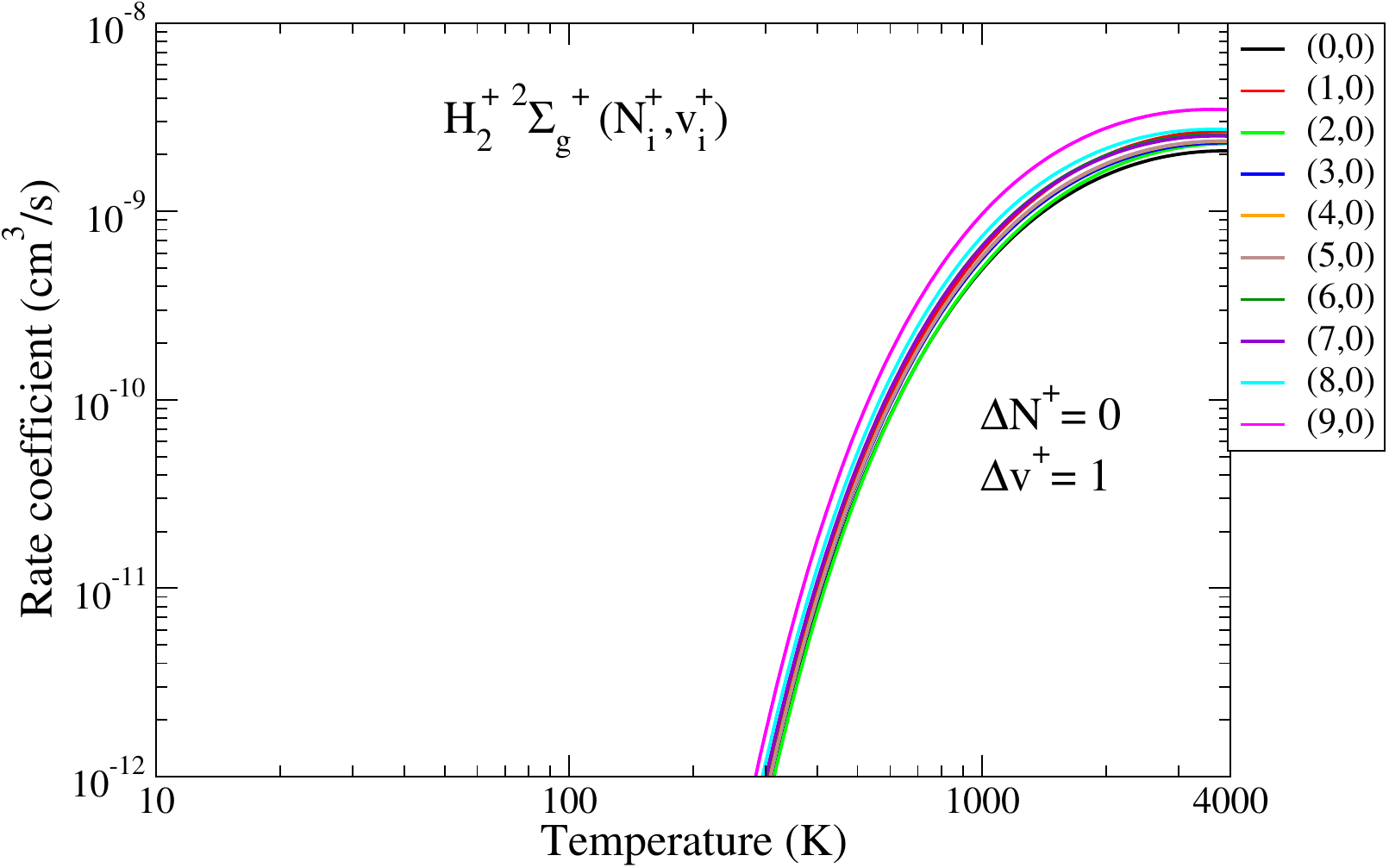}
\includegraphics[width=0.46\linewidth]{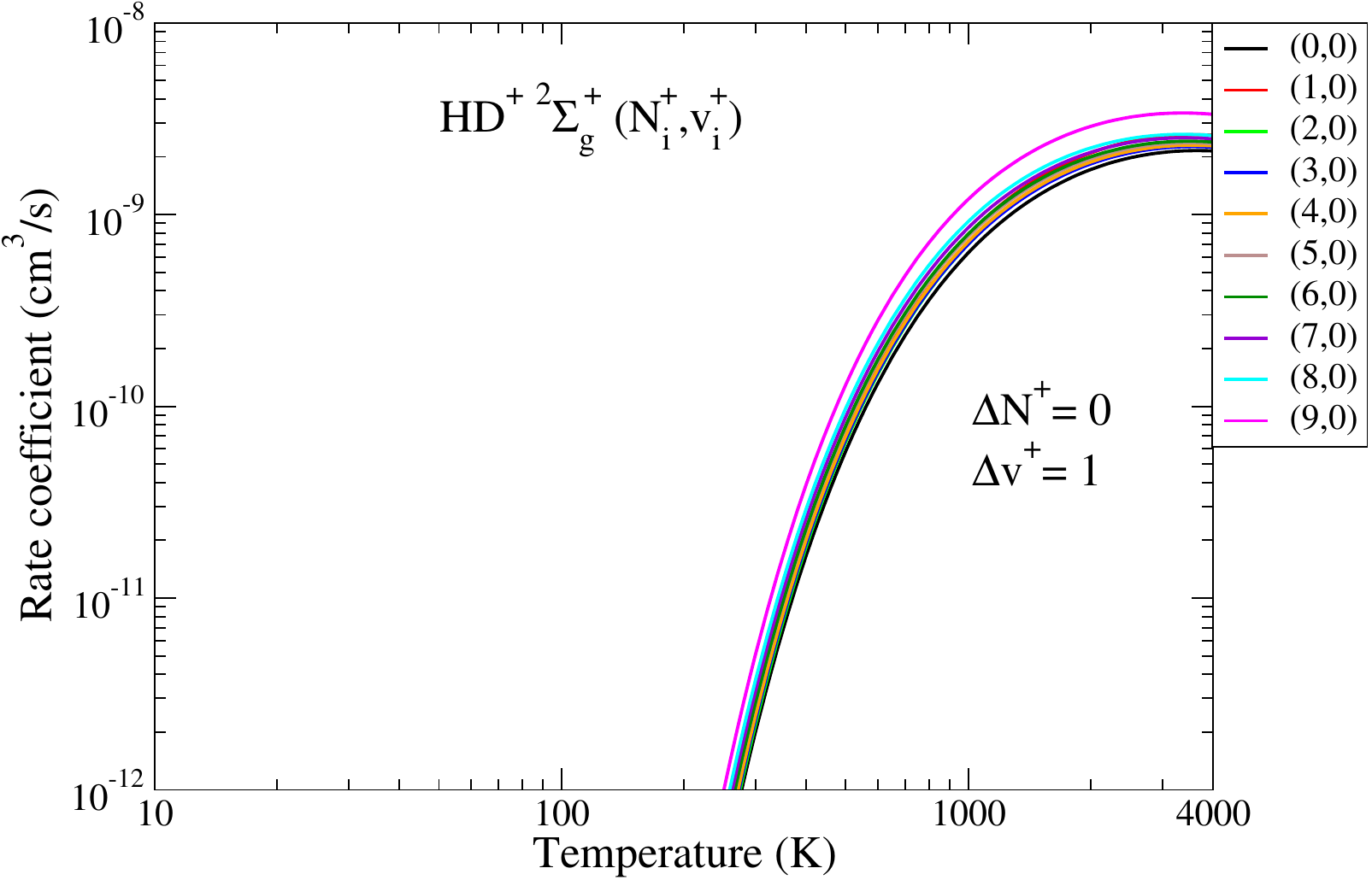}
 \centering
 \caption{{\it Left panel:} 
 electron-impact vibrational excitation ($\Delta v^+=v_f^+-v_i^+=1$), ($\Delta N^+=N_f^+-N_i^+=0$) rate coefficients of H$_2^+$ from its ground $^2\Sigma_g^+$ electronic state and for the first 10 initial ro-vibrational levels $(N_i^+, v_i^+)$, where $v_i^+=0$. 
 {\it Right panels:} same for HD$^+$.}
\label{fig:4}
\end{figure*}

Figures~\ref{fig:1}--\ref{fig:4} display representative samples of RVT cross sections and thermal rate coefficients. More specifically, Figure~\ref{fig:1} displays the cross sections of simultaneous electron-impact rotational excitation ($\Delta N^+=N_f^+-N_i^+=2$) and vibrational de-excitation ($\Delta v^+=v_f^+-v_i^+=-1$) from the ground $^2\Sigma_g^+$ electronic state and initial ro-vibrational levels $(N_i^+, v_i^+)=(0,1)$ and $(N_i^+, v_i^+)=(2,1)$. Figure~\ref{fig:2} displays the corresponding rate coefficients for 10 ro-vibrational levels of the target, ranging from $(0,1)$ to $(9,1)$. Figure~\ref{fig:3} presents rate coefficients for pure rotational excitation ($\Delta N^+ = 2$, $\Delta v^+ = 0$) from levels $(0,0)$ to $(9,0)$, while Figure~\ref{fig:4} shows rate coefficients for pure vibrational excitation ($\Delta N^+ = 0$, $\Delta v^+ = 1$) over the same set of initial levels.

\begin{figure*}
\centering
\includegraphics[width=0.46\linewidth]{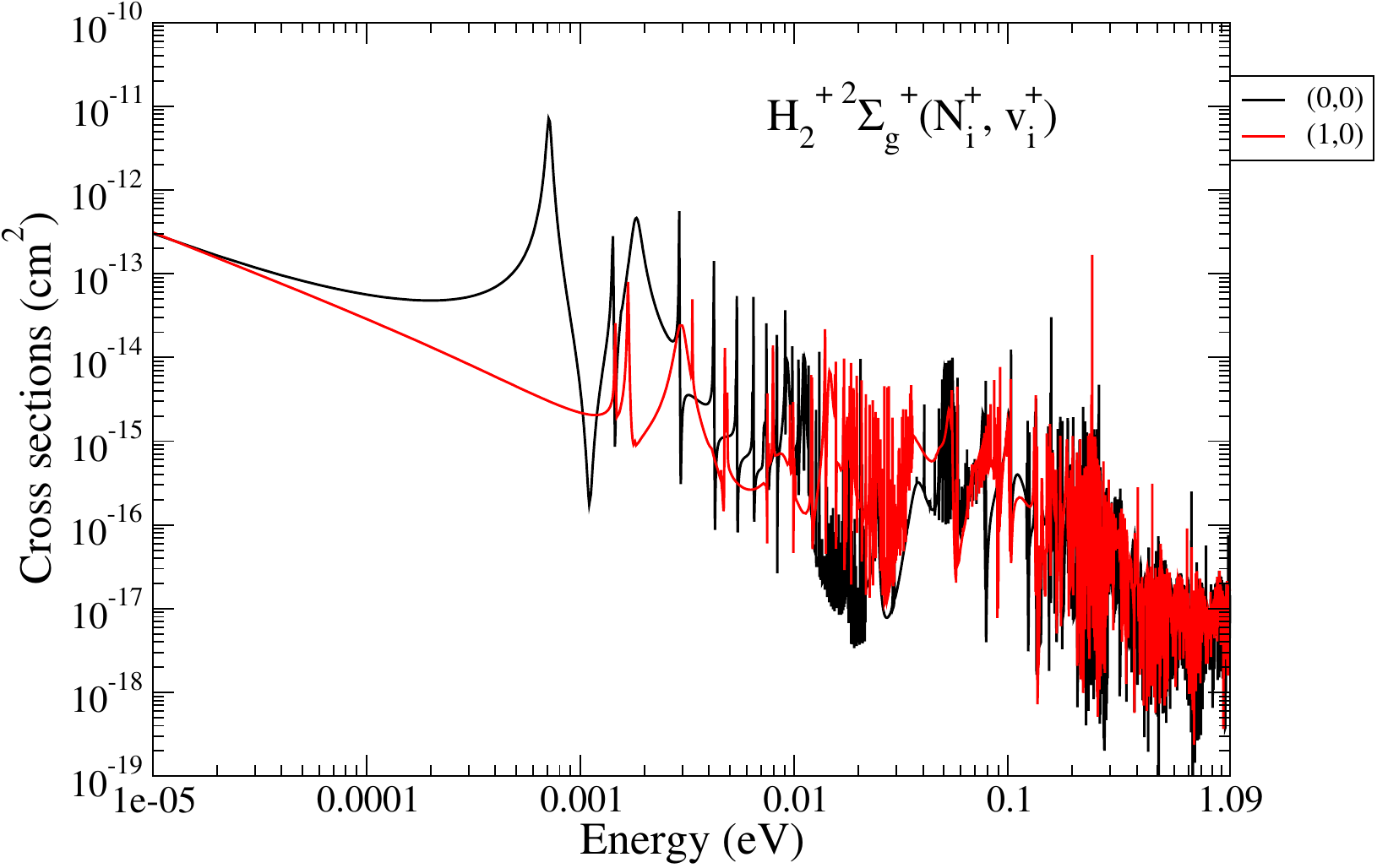}
\includegraphics[width=0.46\linewidth]{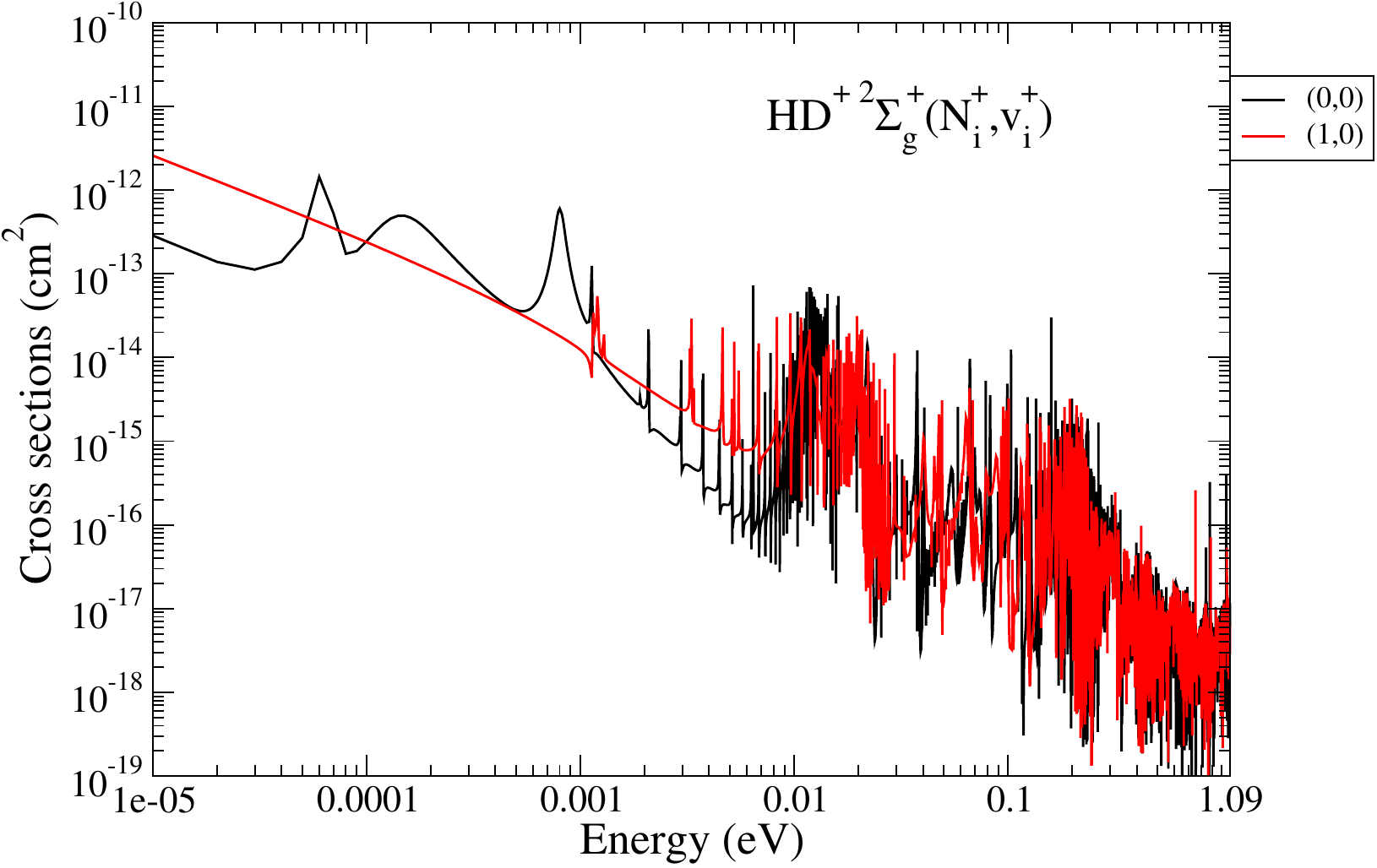}
 \centering
 \caption{
{\it Left panel:} DR cross sections of H$_2^+$ from its ground $^2\Sigma_g^+$ electronic state, and the two lowest initial ro-vibrational levels: $(N_i^+, v_i^+)=(0,0)$, {\it black line}, and $(N_i^+, v_i^+)=(1,0)$, {\it red line}. {\it Right panel:} same for HD$^+$.} 
\label{fig:5}
\end{figure*}

\begin{figure*}
\centering
\includegraphics[width=0.46\linewidth]{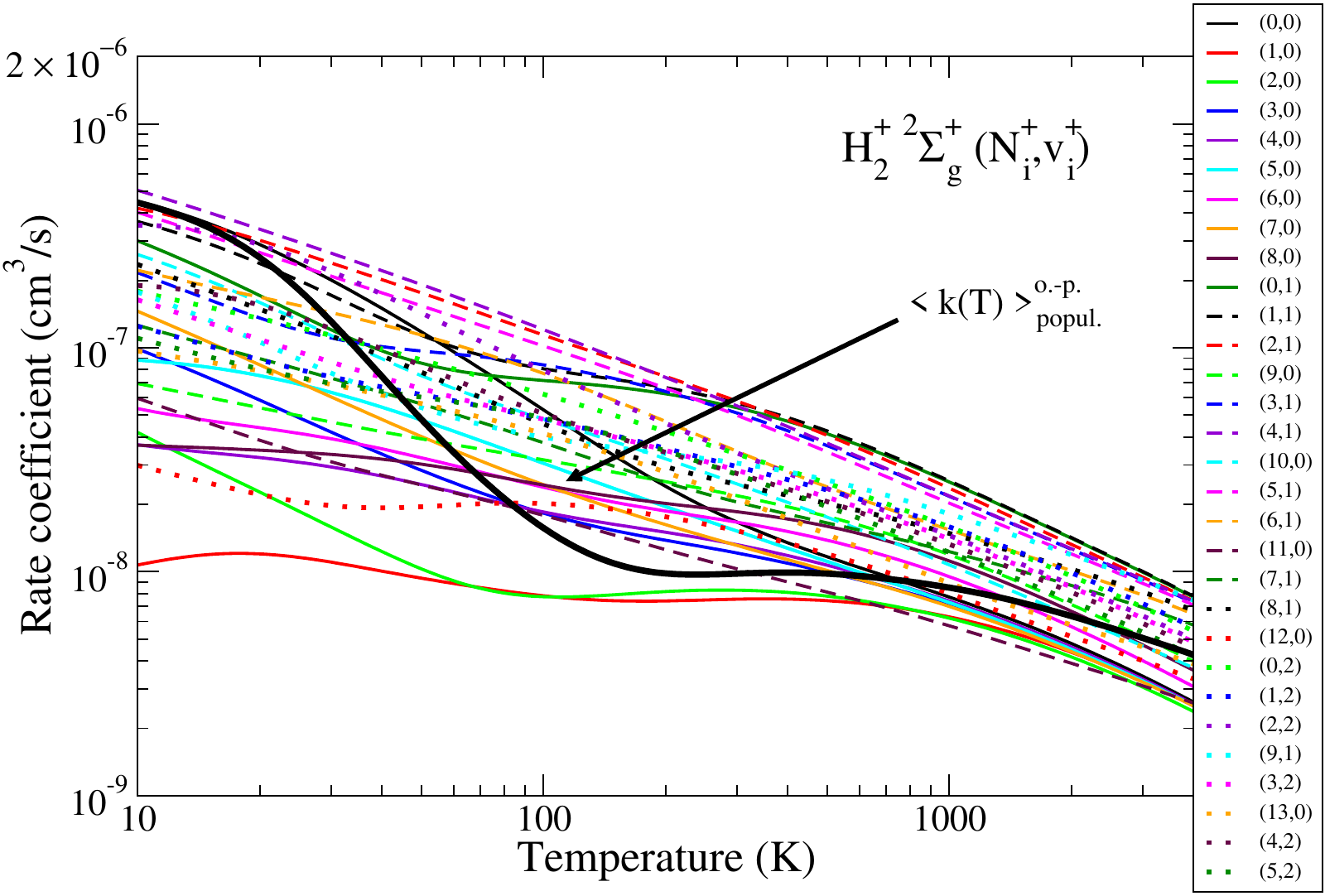}
\includegraphics[width=0.46\linewidth]{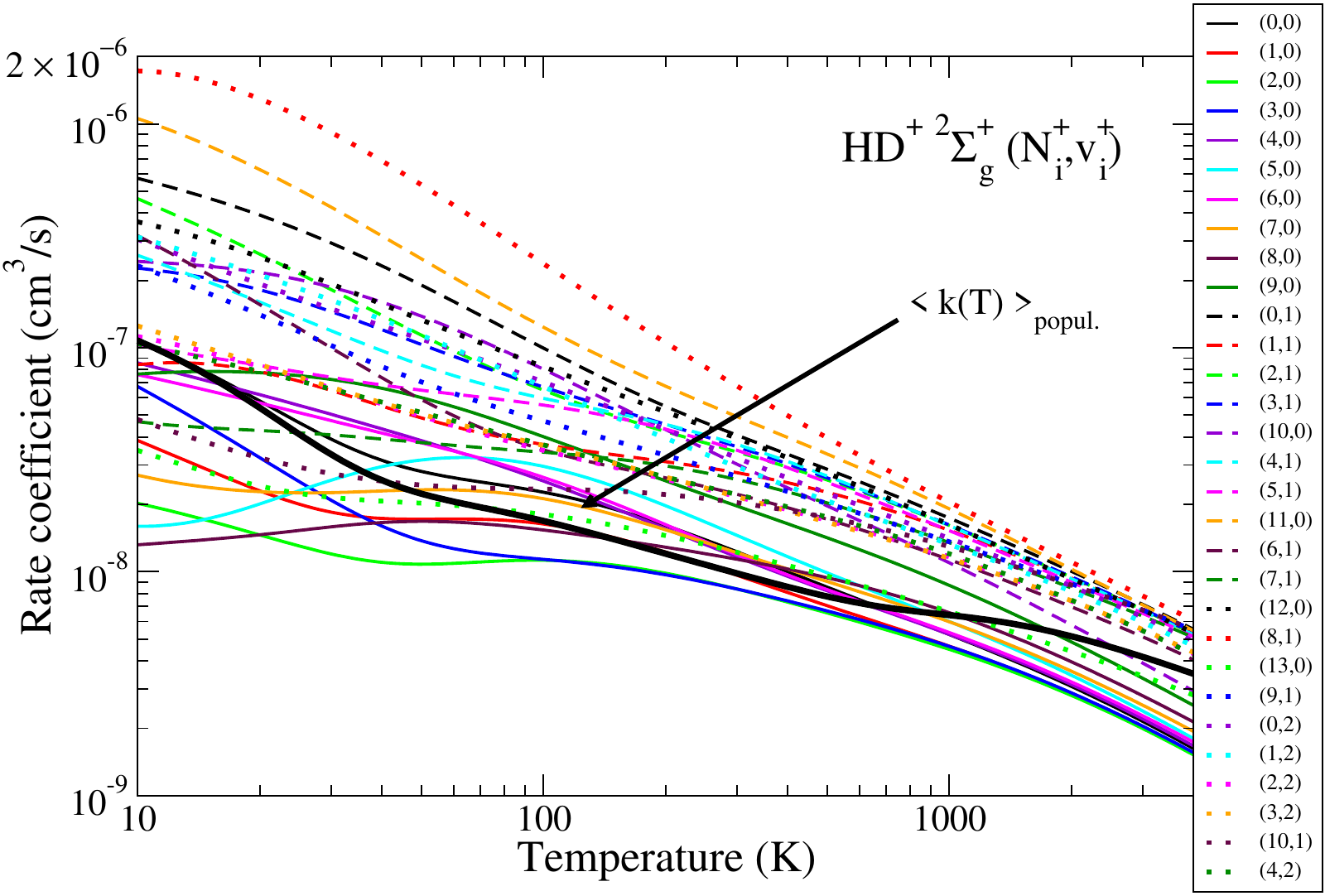}
\caption{{\it Left panel}: thermal rate coefficient for the DR of H$_2^+$ summed over all relevant molecular symmetries up to $T=4000$~K electron temperature.
The {\it thin colored lines} show the rate coefficient for the first 30 ro-vibrational levels ($N_i^+,v_i^+$) listed in the legend, while the {\it thick black line} shows the thermally averaged rate coefficient by considering the temperature-dependent level population distributions and the ortho-para statistical weights. {\it Right panel}: same for HD$^+$ without ortho-para weights.}
\label{fig:6}
\end{figure*}

In parallel, we have computed DR cross sections under the same conditions as those used for the RVT calculations. Figure~\ref{fig:5} presents the results for the two lowest initial levels, $(N_{i}^+, v_{i}^+) = (0,0)$ and $(1,0)$. The corresponding DR rate coefficients, derived from the full set of 30 levels, are shown in Figure~\ref{fig:6} up to an electron temperature of 4000~K, along with the thermally averaged values. The latter incorporates the temperature-dependent population distributions and the ortho–para statistical weights (relevant only for H$_2^+$).

The results have been fitted using the following expression:
\begin{equation}
k(T)=a_{0}\left(\frac{T}{300~{\rm K}}\right)^{a_{1}}e^{-a_2/T}+a_{3}\left(\frac{T}{300~{\rm K}}\right)^{a_{4}}e^{-a_5/T},
\label{arrhenius}
\end{equation}
where the coefficients \{$a_{0}$,$a_{5}$\} have the values listed in Table~\ref{tab:arrhenius} and in Table~\ref{tab:arrhenius2}. In addition, fitting parameters corresponding to the Franck–Condon averaged rate coefficients, for initial vibrational levels $v_i^+ = 0$–2, have been obtained using the population weighting factors reported in \citet{Weijun1993}. While the thermal average is generally valid for applications to diffuse environments, the Franck-Condon average is more appropriate for applications to dense clouds, where the timescale for conversion of 
H$_2^+$ to H$_3^+$ is of the same order as the radiative decay by electric quadrupole transitions of vibrationally-excited H$_2^+$ formed by cosmic-ray ionization of H$_2$.

\begin{table*}
\centering
\caption{Fitting parameters for Eq.~(\ref{arrhenius}), corresponding to the thermally averaged and Franck-Condon averaged rate coefficients for DR of H$_{2}^{+}$, for every rotational level $N_{i}^{+}$, with $v_{i}^{+}=0-2$. Electron temperature is between 1 and 4000 K.} 
\label{tab:arrhenius}
\begin{tabular}{cccccccc}
\hline
\rule{0pt}{3ex} 
$v_{i}^{+}$ & $a_{0}$ & $a_{1}$ & $a_{2}$ & $a_{3}$ & $a_{4}$ & $a_{5}$ & RMS\\
\rule{0pt}{3ex} & ($10^{-8}$~cm$^{3}$~s$^{-1}$) & & (K) & ($10^{-8}$~cm$^{3}$~s$^{-1}$) & & (K) & \\
	    \hline
     \rule{0pt}{3ex}0 & 3.94566 & $-0.890737$ & 5.6937 & $-3.29071$ & $-1.36802$ & 55.7766 & 0.089366\\
      1 & 4.86941 & $-0.542306$ & $-0.142774$ & $-9.51705 \times 10^{-2}$ & -2.41121 & 59.6703 & 0.263542\\
      2 & 2.83337 & $-0.542742$ & 0.30275 & $5.94191 \times 10^{-10}$  & $-10.1195$ & 139.231 & 0.143254\\
      0-2 (Thermal total average) & 3.94566 & $-0.890737$ & 5.6937 & $-3.29071$ & $-1.36802$ & 55.7766 & 0.089366\\
      0-2 (Franck-Condon average) & 2.88675 & $-0.572184$ & 0.12744 & $7.23078 \times 10^{-3}$ & $-2.6078$ & 17.2876 & 0.108678\\
\hline
\end{tabular}
\end{table*}

\begin{table*}
\centering
\caption{Fitting parameters for Eq.~(\ref{arrhenius}), corresponding to the thermally averaged rate coefficients for DR of HD$^{+}$, for every rotational level $N_{i}^{+}$, with $v_{i}^{+}=0-2$. Electron temperature is between 1 and 4000 K.}
\label{tab:arrhenius2}
\begin{tabular}{cccccccc}
\hline
\rule{0pt}{3ex} 
$v_{i}^{+}$ & $a_{0}$ & $a_{1}$ & $a_{2}$ & $a_{3}$ & $a_{4}$ & $a_{5}$ & RMS\\
\rule{0pt}{3ex} & ($10^{-8}$~cm$^{3}$~s$^{-1}$) & & (K) & ($10^{-8}$~cm$^{3}$~s$^{-1}$) & & (K) & \\
	    \hline
     \rule{0pt}{3ex}0 & $0.85269$ & $-0.3476$ & -1.08852 & $5.00446 \times 10^{-2}$ & $-1.63551$ & 6.18157 & 0.0965788\\
      1 & 2.58153 & $-0.840242$ & 0.615672 & $9.247 \times 10^{-3}$ & $-3.21729$ & 34.9419 & 0.277376\\
      2 & 2.58878 & $-0.739446$ & 0.475762 & $4.25466 \times 10^{-3}$ & $-4.70379$ & 151.291 & 0.0852925\\
      0-2 (Thermal total average) & 1.66967 & $-0.58484$ & 0.93083 & -0.631148 & $-1.19178$ & 38.9742 & 0.0898261\\
      \hline
	\end{tabular}
\end{table*}

\begin{figure*}
\centering
\includegraphics[width=0.45\textwidth]{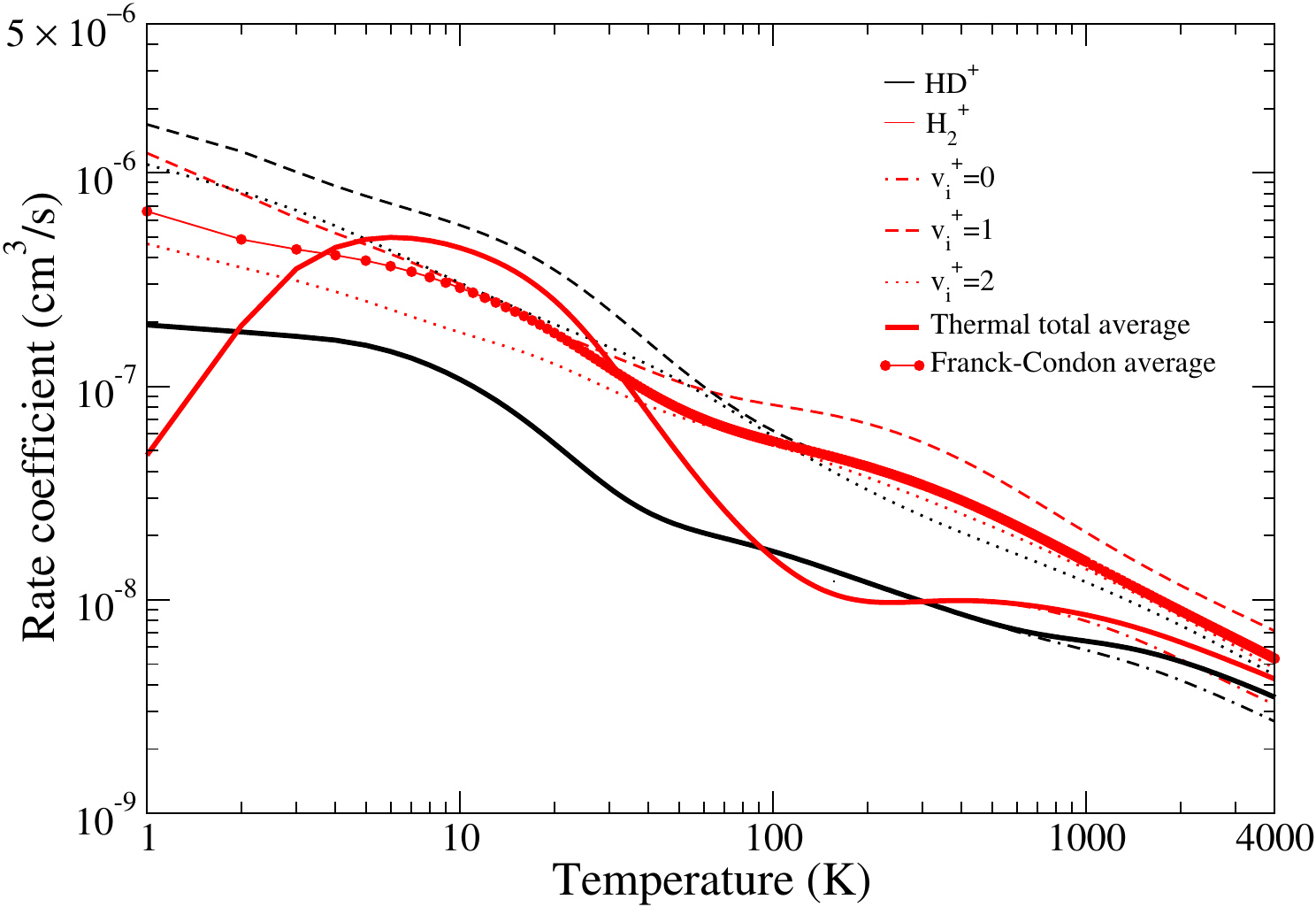}
\caption{
Thermally averaged and Franck-Condon averaged rate coefficients for DR of H$_2^+$ and HD$^+$ for every rotational level $N_i^+$, with $v_i^+=0-2$ ({\it red and black lines}) using the fitting parameters given in tables \ref{tab:arrhenius} and \ref{tab:arrhenius2}. 
Electron temperature is between 1 and 4000~K.}
\label{fig:7}
\end{figure*}

Figure~\ref{fig:7} presents the fitted curves for the thermally averaged and Franck–Condon averaged DR rate coefficients of H$_2^+$ (red curves), along with the fits of the thermally averaged DR rate coefficients of HD$^+$ (black curves), for initial vibrational levels $v_i^+ = 0$, $1$, and $2$, up to an electron temperature of 4000~K.

\section{Astrophysical applications}
\label{sec:app}

\begin{figure*}
\centering
\includegraphics[width=0.40\linewidth]{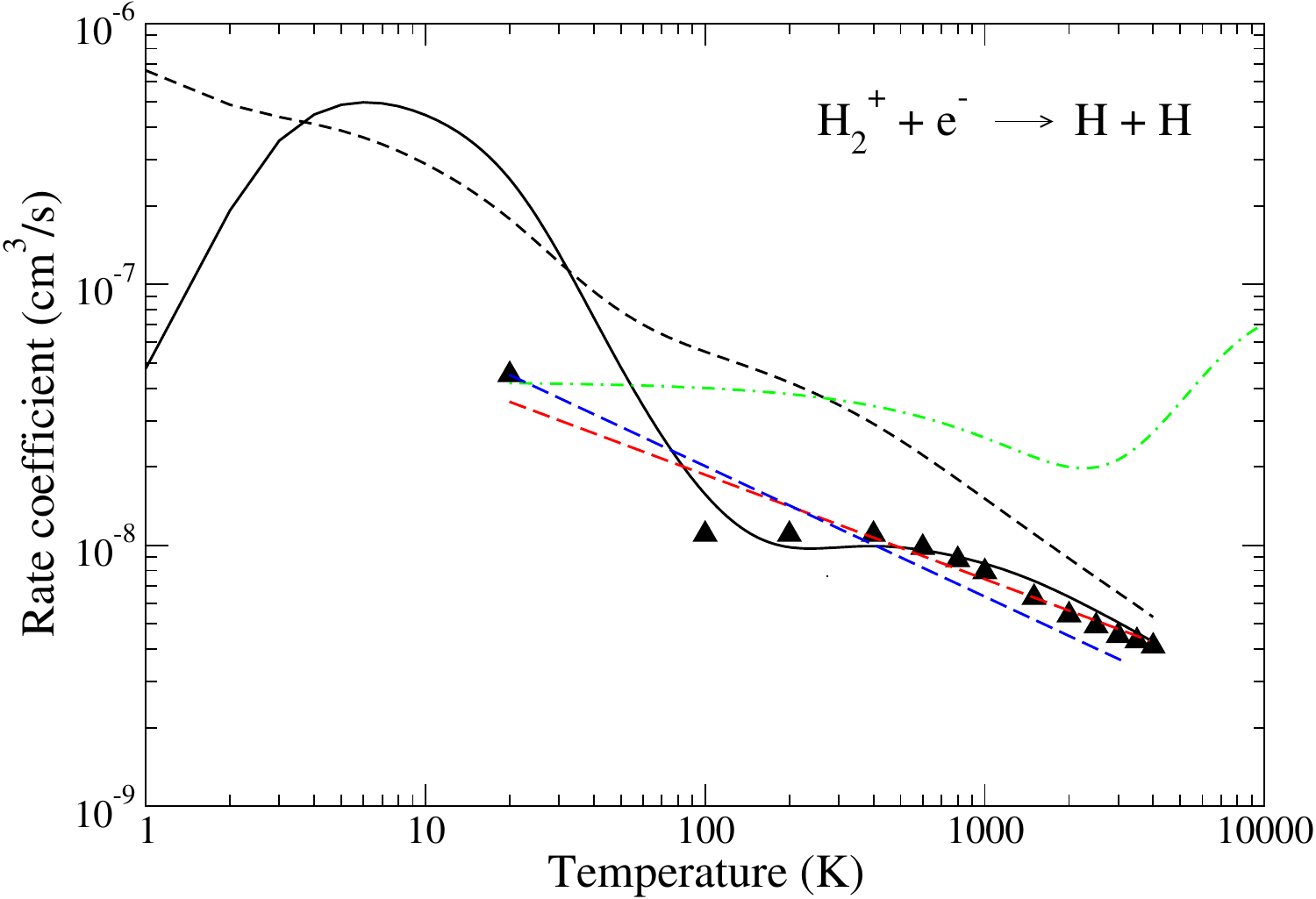}
\includegraphics[width=0.40\linewidth]{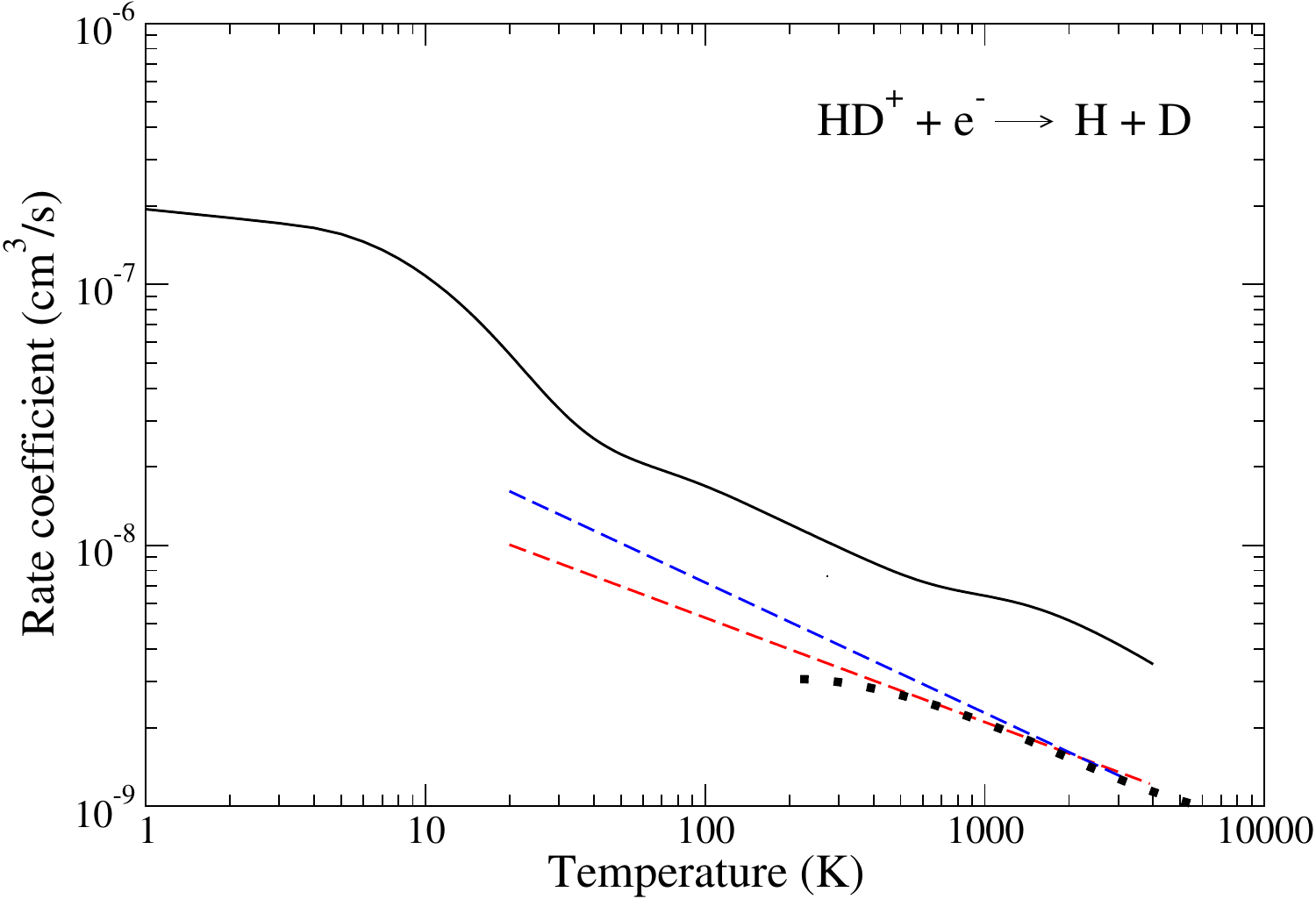}
\caption{Comparison of thermal rate coefficients for the DR of H$_2^+$ and HD$^+$. {\it Left panel}:
DR of H$_2^+$. {\it Blue} and {\it red dashed lines:} fits by \protect\cite{gp98} and 
\protect\cite{sld98}, respectively, of the thermal rate coefficient computed by \protect\cite{schneider1994} ({\it triangles}); {\it green dot-dashed line:}
fit of the rate computed by \protect\cite{coppola2011} from 
the cross sections by \protect\cite{takagi2002}; {\it black line:} this work. The black dashed line corresponds to our calculated Frank-Condon rate coefficient.
{\it Right panel}: Same for HD$^+$. {\it Blue} and {\it red dashed lines:} extrapolations by \protect\cite{gp98} and 
\protect\cite{sld98}, respectively, of the thermal rate coefficient computed from the experimental cross section by \protect\cite{str95} ({\it dots}\/); {\it black line:} this work.}
\label{fig:8}
\end{figure*}

Figure~\ref{fig:8} shows a comparison of the thermal rate coefficients of H$_2^+$ and HD$^+$ adopted in models of the chemistry of primordial gas, compared to our results. The calculations presented in this paper (black curves) resolve considerable discrepancies in previous evaluations of the DR rate of H$_2^+$ and HD$^+$, and extend the results to temperatures below $\sim 100$~K, where all available estimates of the DR rate significantly underestimate its importance. 

In the early Universe, DR of H$_2^+$ and HD$^+$ competes with photodissociation (at $z\gtrsim 300$) and charge exchange with H (at $z\lesssim 300$) in the destruction of these species. 
At low redshifts ($z\lesssim 100$) the significant increase of DR at low temperatures evidenced in the current work implies a larger effect of DR with respect to previous results obtained with the DR rate by \cite{schneider1994}: at 
$z \approx 10$--$20$, the contribution of DR of H$_2^+$ and HD$^+$ increases from $\sim 2$\% to $\sim 12$\% and from $\sim 1$\% to $\sim 5$\% of the total destruction rate, respectively. However, because of the low electron fraction, the destruction of both cations at low redshift remains dominated by charge-exchange reactions with the more abundant H atoms. 

At high redshift, the intense cosmic radiation background is by far the dominant destruction process for all molecular ions, and DR plays no role. This is not the case when the radiation field is produced by a point source, as in planetary nebulae \citep{black1978} and H{\sc ii} regions \citep{aleman2004}, because of the geometric dilution of radiation with distance from the source. In addition, DR in nebulae is also favored by the high abundance of free electrons. In this situation, the abundance of H$_2^+$ is basically controlled by the balance of formation by radiative association of H and H$^+$ (with rate $k_{\rm ra}$) and destruction by DR (with rate $k_{\rm dr}$), and it is given by H$_2^+$/H = $k_{\rm ra}/k_{\rm rd}$ \citep[see, e.g.][]{cecchi1993}. 

\begin{figure*}
\centering
\includegraphics[width=0.40\linewidth]{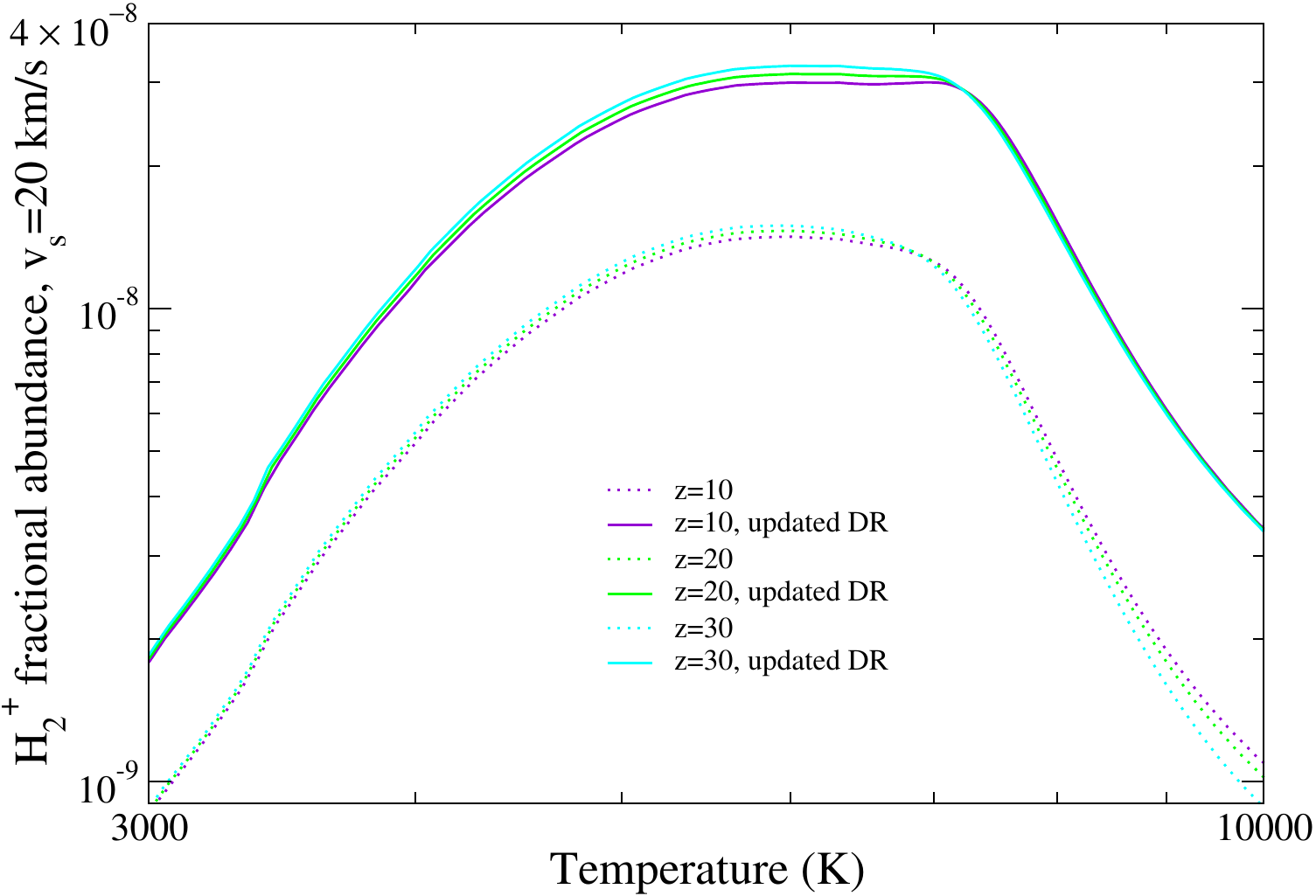}
\includegraphics[width=0.40\linewidth]{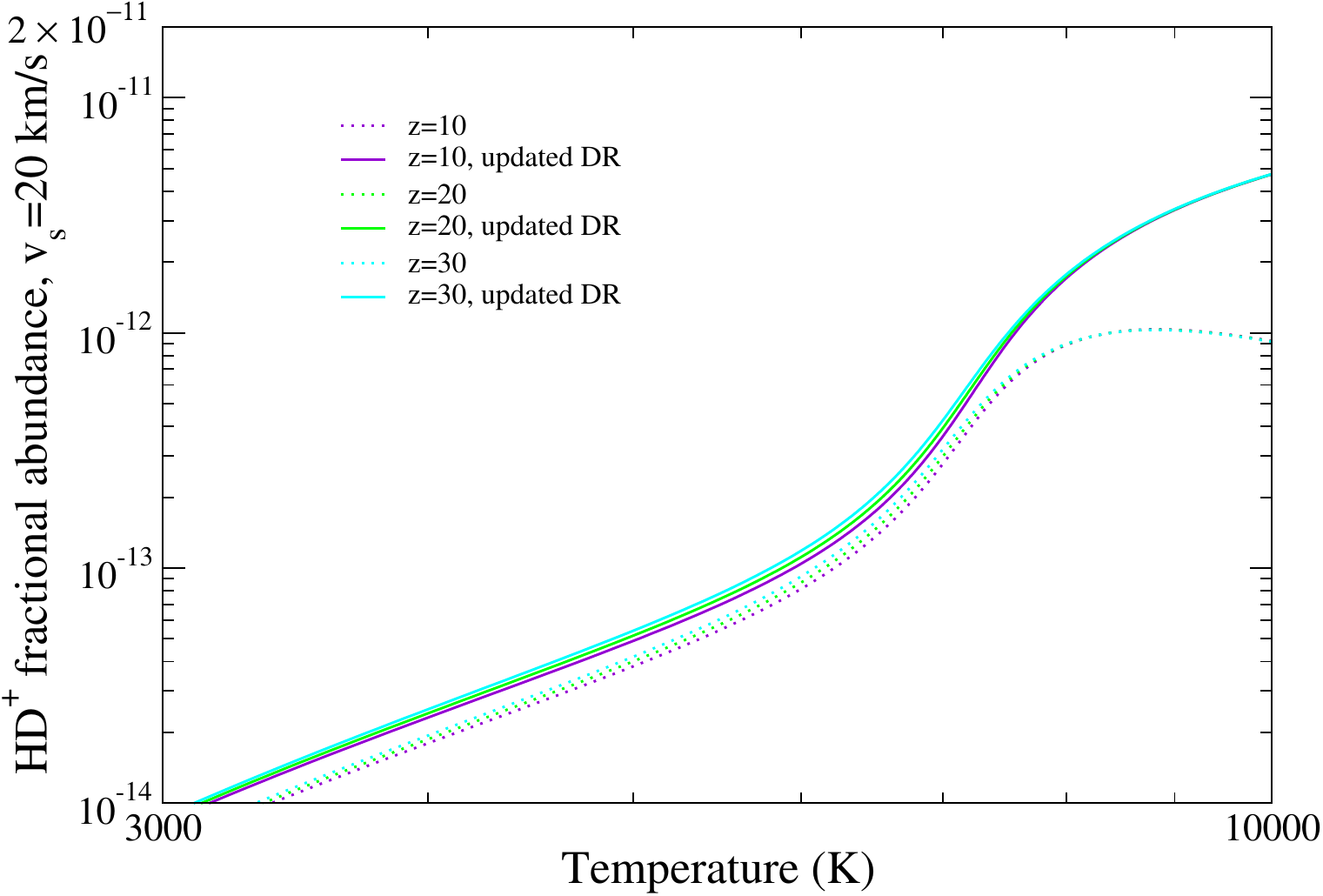}
\caption{Fractional abundances of H$_2^+$ ({\it left panel}\/) and HD$^+$ ({\it right panel}\/) for a shock wave propagating through a gas of primordial composition at redshift $z=10$, 20, and 30 and at a speed $v_s=20$~km~s$^{-1}$. The dotted curves show the fractional abundances obtained adopting the DR reaction rates of \protect\cite{coppola2011} and \protect\cite{gp98}, respectively.}
\label{fig:9}
\end{figure*}

Shock waves could also lead to high H$_2^+$ and electron abundances. For this reason, accurate and state-to-state resolved DR reaction rates are needed \citep{shapiro1987, takagi2002, vasiliev2006, coppola2016}. In particular, as shown in Figure~\ref{fig:9}, the updated DR reaction rates in the chemical evolution of shock waves through gas at high redshift, lead to changes in the fractional abundances of H$_2^+$ and HD$^+$. Specifically, these simulations have been carried out at $z=10$, $20$, $30$ and for a shock speed $v_s=20$~km~s$^{-1}$, compatible with typical primordial supernova explosions. Differences of up to a factor of $\sim 2$ in the case of H$_2^+$ and a factor of $\sim 4$ in the case of HD$^+$ are found in the temperature range between 3000~K and $10^4$~K (for this particular example, the fitting formula given by Eq.~\ref{arrhenius} has also been adopted at temperatures higher than the upper limit of the calculated rate, $T=4000$~K). 

\section{Conclusions}
\label{sec:con}

We have computed cross sections and thermally averaged rate coefficients summed for all relevant molecular symmetries from $T=1$~K up to $T=4000$~K electron temperatures considering temperature-dependent level population distributions. These new cross sections and rate coefficients can be adopted to model the kinetics of hydrogen-rich media in which the ionization fraction is significant. A primary field of application is planetary atmospheres and astrophysical plasmas in general, either in conditions typical of the early Universe (absence of metals and dust grains) or in the present-day partially ionized interstellar medium (as, for example, in planetary nebulae and H{\sc ii} regions). In all these environments, electron-impact induced ro-vibrational transitions are the dominant excitation process of H$_2^+$ and HD$^+$, and dissociative recombination is one of their main destruction channels. Our main findings are summarized as follows:

   \begin{enumerate}
      \item With the new cross sections, significant differences with previously adopted thermal rate coefficients of dissociative recombination are found (see Sect.~\ref{sec:app} and Fig.~\ref{fig:9}).
      \item The new rates produce substantial differences with respect to earlier results when introduced in astrochemical models, for example in models for the shock-induced chemistry of hydrogen/deuterium in a zero-metal gas (see Fig.~\ref{fig:8}).
   \end{enumerate}

\section*{Data availability}
\label{sec:data}

The data supporting this article will be provided by the corresponding author.

\begin{acknowledgements}
      The authors acknowledge support from Agence Nationale de la Recherche {\it via} the project MONA, Centre National de la Recherche Scientifique {\it via} the GdR TheMS, PCMI program of INSU (ColEM project, co-funded by CEA and CNES), PHC program Galil\'ee between France and Italy, and DYMCOM project, F\'ed\'eration de Recherche Fusion par Confinement Magn\'etique (CNRS, CEA and Eurofusion), La R\'egion Normandie, FEDER and LabEx EMC$^3$ {\it via} the projects Bioengine, EMoPlaF, CO$_2$-VIRIDIS and PTOLEMEE, COMUE Normandie Universit\'e, the Institute for Energy, Propulsion and Environment (FR-IEPE), International Atomic Energy Agency (IAEA) via de project CRP: ``The Formation and Properties of Molecules in Edge Plasmas'' and the European Union  via COST (European Cooperation in Science and Technology) actions TUMIEE (CA17126), MW-GAIA (CA18104), MD-GAS (CA18212), COSY (CA21101),  PLANET(CA22133), PROBONO (CA21128), PhoBioS (CA21159), DAEMON (CA22154), DYNALIFE (CA21169), and ERASMUS-plus conventions between Universit\'e Le Havre Normandie and Politehnica University Timi\c{s}oara, West University Timi\c{s}oara and University College London. IFS acknowledges the support from the French PEPR SPLEEN consortium, via the project PLASMA-N-ACT. NP is grateful for the support of the Romanian Ministry of Research and Innovation, project no. 10PFE/16.10.2018, PERFORM-TECH-UPT. JZsM thanks the financial support of the National Research, Development and Innovation Fund of Hungary, under the FK 19 and Advanced-24 funding schemes with project no.~FK 132989 and 151196. DG acknowledges support by INAF grant PACIFISM.
\end{acknowledgements}

%
%

\end{document}